# Complex Networks, Communities and Clustering: A survey

Biswajit Saha[1], Amitabha Mandal[1], Soumendu Bikas Tripathy[2], Debaprasad Mukherjee[1]

Department of Computer Science & Engineering and Department of Information Technology, Dr. B. C. Roy Engineering College, Durgapur [1], West Bengal, India

Department of Computer Science & Engineering, Durgapur Institute of Advanced Technology & Management, Durgapur [2],

frombiswajit@rediffmail.com, amitabha82@gmail.com, sbtripathy@gmail.com, mdebaprasad@gmail.com

## Abstract

This paper is an extensive survey of literature on complex network communities and clustering. Complex networks describe a widespread variety of systems in nature and society especially systems composed by a large number of highly interconnected dynamical entities. Complex networks like real networks can also have community structure. There are several types of methods and algorithms for detection and identification of communities in complex networks. Several complex networks have the property of clustering or network transitivity. Some of the important concepts in the field of complex networks are small-world and scale-free networks, evolving networks, the relationship between topology and the network's robustness, degree distributions, clustering, network correlations, random graph models, models of network growth, dynamical processes on networks, etc. Some current areas of research on complex network communities are those on community evolution, overlapping communities, communities in directed networks, community characterization and interpretation, etc. Many of the algorithms or methods proposed for network community detection through clustering are  modified versions of or inspired from the concepts of minimum-cut based algorithms, hierarchical connectivity  based algorithms, the original Girvan–Newman algorithm, concepts of modularity maximization, algorithms utilizing metrics from information and coding theory, and clique based algorithms.

## 1. Introduction

### 1.1 Complex Networks

Complex networks describe an extensive variety of systems in nature and society esp. systems composed by a large number of highly interconnected dynamical entities. The Internet, social networks, business networks, large circuits, network of chemicals linked by chemical reactions, transportation networks, power networks, network of citations of documents/ web pages, etc, are some of the popularly cited examples of complex networks. Two of the many important questions in this field are: a) are there any unifying principles underlying the topology complex networks, and b) from the perspective of nonlinearity, how colossal networks of interacting and/or communicating dynamical systems will behave collectively, given their individual dynamics and coupling architecture. One of the intuitive approaches to capture the global properties of such complex systems is to model them as graphs, where nodes represent the dynamical units, and links/edges represent the interactions between the nodes. It has now been widely recognized that the topology and evolution of real world complex networks are controlled by various organizing principles of topology and dynamics. Researchers have been addressing structural and topological issues of complex networks, e.g. a) characterization of the complex interconnection architectures, to comprehend the unifying principles that are the foundations of real-world complex networks, and b) constructing models to simulate the growth and replicate the structural properties of these complex networks. Researchers have also addressed the complex networks' dynamics, e.g. characterization of the collective behaviour of large ensembles of dynamical systems that interact through complex interconnections topology. Some of the important concepts in the field of complex networks are small-world and scale-free networks, evolving networks, the relationship between topology and the network's robustness, degree distributions, clustering, network correlations, random graph models, models of network growth, dynamical processes on networks, etc. [1-4]

### 1.2 Network Communities

Community structures are quite common in real networks. Complex networks have community structure if the nodes of the networks can be grouped into sets of nodes where each set of nodes is densely connected, at least, internally. For non-overlapping network communities, the complex network should split into groups of nodes with dense connections internally, but sparser connections between the groups. Overlapping network communities are also possible, and they are found in many real world complex networks e.g. social networks. An alternative way of expressing the network community concept, inclusive of the overlapping and non-overlapping issues, is that, pairs of nodes are more likely to be connected if they are both members of the same community(ies), and less likely to be connected if they are not members of the same communities. The communities can themselves also join together to form meta-communities, and that those meta-communities can join together, and so on in a hierarchical fashion. It is now well accepted that the identification of the community structure of complex networks provides insight into the relationships between network function and topology. There are several types of methods and algorithms for detection and identification of communities in complex networks. But, roughly, these methods can be divided into 4 categories (not exclusive): a) node-centric (algorithms utilizing the information that each node in a group satisfies certain properties) b) group-centric (algorithms utilizing the connections within a group as a whole, i.e. the group as a single entity satisfies certain properties without zooming into node-level), c) network-centric (algorithms utilizing the process of partitioning of the whole network into several disjoint sets), and d) hierarchy-centric (algorithms utilizing the processes of construction of hierarchical structures of communities). There are other categories of algorithms/methods of community detection in complex networks, but they are not being discussed here. The optimal methods to detect network communities vary depending on applications, complex networks, computational resources etc. Some current lines of research on complex network communities are those on community evolution, overlapping communities, communities in directed networks, community characterization and interpretation, etc.[5-8].

**1.3 Clustering**

Several complex networks have the property of clustering or network transitivity. It is the property that two vertices of the network that are both adjacent of the same third vertex have an increased probability of also being adjacent of one another. Clustering of network communities is the task of grouping a set of nodes in such a way that nodes in the same group (cluster) are more similar, in some measure(s), to each other, than to those in other groups (clusters), based on the above mentioned adjacency property. The choice or design of clustering algorithms and their corresponding controlling functions, e.g. distance measure/norm, threshold/cut-offs for similarity/dissimilarity measures, number of final expected stable clusters, degrees of randomization, criteria for cluster partitioning or merging, etc., depend on the complex networks being clustered and the nature and function of the communities to be identified in them. Clustering of network communities can also be formulated as a multi-objective optimization problem. The differences in the cluster/community models and their properties reflect the differences between the various clustering and network community detection algorithms. Some popular cluster/community models are a) connectivity models, e.g. hierarchical clustering algorithms create models utilizing distance-based interconnections between nodes or already existing communities, b) centroid models, e.g. models those characterize clusters/ network communities by mean vectors computed from set of properties of sets of nodes, c) distribution models based on statistical distributions of the values features of the nodes in multidimensional space, d) density models, based on density or sparseness of connections/ regions in data space, and a very important category e) subspace models e.g. in bi clustering/co-clustering where clusters/network communities are modelled with respect to both cluster members and pertinent set of features of the nodes, g) graph-based models e.g models/ algorithms based on quasi-cliques- subset of nodes in a graph/complex network, where a significant fraction of node pairs in the subset are connected by edges, considered as cluster/ community seeds for building up larger clusters or communities. Many of the algorithms or methods proposed for network community detection through clustering are modified versions of or inspired from the concepts of minimum-cut based algorithms, hierarchical connectivity based algorithms, the original Girvan–Newman algorithm, concepts of modularity maximization, algorithms utilizing metrics from information and coding theory, and clique based algorithms [9-12].

## 2. Classification of Community Detection Clustering Algorithms in Complex Networks

Community detection clustering algorithms in complex networks can be classified in the following ways:

### 2.1 Hierarchy based algorithms

Hierarchy based algorithms is one of the main type of community detection clustering algorithms in complex networks. Silva et al. [13] utilized the concept of topological orders among input data represented as graph, and developed an algorithm to obtain clusters in different scales. The algorithm consisted of initially the network construction from input data, and secondly, the hierarchical partitioning of the formed network. The algorithm, although being completely free of the computation of the physical distances among input data, was found to constantly produce a connected graph with heavily linked nodes within a community and sporadically linked nodes among different communities. The authors applied their algorithm to the problem of pixel clustering.

Zhang et al.[14] proposed a hierarchical clustering approach, based on the graph diffusion kernels of networks, to reveal the community organization of different levels of complex networks. They verified the method on some networks with well-known community structures, and found that the algorithm is an effective one.

Guang Xu et al.[15] proposed a new algorithm, called Latent Community Discovery, for community detection in complex social networks. Specifically, their algorithm divides the core actors, based on a hierarchical probabilistic model and a statistical topic model, which are normalized by the network arrangement in data. Their algorithm is inspired from the Pareto Principle, which accounts for the uneven existence of two different types of network actors, esp. the core actors who typically occupy only a small share of the nodes, but have a large influence on the complex network. They tested their algorithm on three large social networks, and found its performance competitive to the existing popular algorithms for this category of problems.

de Oliveira et al. [16] propose a clustering algorithm based on graph theoretic representations and community discovery in complex networks. Initially, they represent the input data as a network, and then divide the network into sub-networks to create data groups. In the first stage, each of the nodes has a randomly assigned initial angle. This initial angle is gradually modified according to agreement with the neighbours' angle. Ultimately the network reaches a steady state. In this state the nodes in the same cluster have comparable angles. This process is repeated and results in a hierarchical and graded divisive clustering. Simulations by the authors demonstrate that this algorithm has the potential to find clusters in different forms, compactness and proportions. The algorithm also has the capability to generate clusters with diverse refinement grades. Furthermore, the proposed algorithm is also robust and efficient.

## 2.2 Information theoretic algorithms

Information theoretic algorithms are another major type of community detection clustering algorithms in complex networks. Cravino et al. [17] employed the overlapping community arrangement of a linkage of tag/labels to improve text clustering. Based on a small data set of news clips/ excerpts, the authors construct a network of co-occurrence of user-defined labels of metadata fields in news excerpts. They describe a weighted cosine similarity closeness measure, which takes into account both the excerpt vectors and the tag vectors. Thereafter, they compute the tag weight using the correlated tags that are existing in the discovered community, and then use the ensuing vectors together with a novel distance metric, to identify socially biased document clusters.

Yang et al. [18] proposed a unsupervised graphical clustering algorithm for finding community in complex networks by determining the dissimilarity between nodes and incorporating them into a dissimilarity distance matrix, which when sorted according to the scores, becomes equivalent to an intensity image, and the clusters are indicated by dark blocks of pixels along the main diagonal.

It is now acknowledged that trust in electronic commerce has become one of the most significant concerns in online applications, with consumers searching for the best trustworthy of goods and service providers and looking for ways of confirmation of which service providers are the most trusted. Zhang et al. [19] have studied the critical problem of trust network and trust community clustering for, the analysis of the users most trusted relationship, for electronic commerce applications. In their model, the nodes represent the various subjects involved in the trust, the connections denote relationships, and the weight of the links indicates the strength of the relationships. Initially, the algorithm constructs a trust network having the weight value of links. Subsequently, the clustering properties of the relationship according to the weights and the path lengths are analysed. Finally, the algorithm categorizes the most trusted subjects for a user to the same cluster. Two metrics, i.e. direct trust information degree and global trust information degree are utilized to assess trust relationships among the subjects. This principle also gives an efficient shortest path algorithm to construct trust networks. All the above information generated are incorporated into the clustering algorithm based on coefficient and path length, for e-commerce trust network community.

Zhang and Zhong [20] dealt with this problem considering the aspects of small world nature of trust communities and the metric local trust recommendation degree.

Piccardi et al. [21] analysed the issue of clustering of financial time series, based on the network community analysis methodology i.e. the partitioning the nodes of a network. A network with n nodes is associated to the set of n time series, and the weight of the link which quantifies the similarity between the two corresponding time series, is defined according to a metric based on symbolic time series analysis. Thereafter, probing for network clusters leads to identification of groups of nodes (i.e. Time series) with strong similarity. The authors verify the algorithm on US and Italian stock exchange time series data and the steadiness of the clusters over time is seen to be satisfactory and better than those achieved using the minimal spanning tree and the hierarchical tree based algorithms.

Piccardi and Calatroni [22] studied the same problem in its full generality, according to the same methodology, in a previous work, and found satisfactory results.

Though community detection in social networks is usually based on graph clustering employing the structural information i.e. linkage structure or node topology, for group identification, but Huang and Yang [23] used the semantic information present in the posts of social media to find hidden communities in these media. They incorporated the assumption that content issued by users may express relations between users/entities. This method is suitable for detecting communities in networks which continuously evolve e.g. social networks.

Liu et al. [24] applied the concept of network community clustering to an important biomedical problem known as functional analysis of protein cavities. It is known that functions of a protein are chiefly determined by its structure, and surface cavities i.e. pockets or clefts, are generally considered as possibly active sites where the protein carries out the functions. The authors proposed a feasible solution to the problem of functional assignment by protein cavity clustering i.e. geometrical clustering based on geometrical community structure of pocket similarity networks. Firstly, they introduce a pocket similarity network to methodically describe structural correspondence among pockets. The pockets are connected if they have structural similarity beyond a certain threshold. Thereafter, the surface pockets are clustered into structurally related pocket groups via a graded process. The authors then reference these these small pocket groups as structural patterns which represent similar functions in different proteins. Their experimental results show identified pocket groups are biologically meaningful in terms of their functional features.

Rui and FengMing [25] proposed an algorithm to construct a distributed trust network in information sharing and exchange channels e.g. instant messengers, file-sharing tools etc. But, it is well known that the formation of the communities is a self-organizing and evolving autonomous phenomenon, being regulated primarily through internal and member dynamics of the community. For example, users join communities based on their interests. These kinds of network communities are highly vulnerable to the spreading of malicious software, pilfering of user's information, attack by malicious users. Thus, to form safe and reliable communities, the authors present an algorithm by incorporating the trust value of nodes computed based on their past behaviours, and constructing the communities based of similarity in trust values.

XIE et al. [26] propose a new algorithm for detecting community structures in a weighted complex networks. The method constructs a weighted complex network with respect to the similarity between document pairs calculated by the cosine function, and then the algorithm searches for the dense sets, applies it to cluster text documents represented by the vector space model.

ZHANG et al. [27] dealt with the issue of trust in e-commerce based on social networks using the metric of trust information degree based on mutual information between subjects. They incorporated their previously developed metrics direct trust information degree and global trust information degree to build trust relations among subjects. Clustering coefficient and global trust information degree were adopted to construct trust community.

Guan-yu [28] has developed an algorithm (named Mapping Vertex into Vector algorithm) which converts all vertices in network into vectors, and find the communities in large scale complex networks through clustering, based on the similarity between these vertex vectors.

You-yuan et al. [29] addressed the problem of web services clustering with the help of the detection of community structure in complex networks. They did this by proposing an algorithm in which the words are denoted by nodes, and the edge-weights of the network are computed from the words' co-appearances. The authors applied the Newmann's algorithm to this network and extracted the clusters of words. Thus service clustering was achieved by using the relationship between the words and the services, and that too with acceptable levels of precision and accuracy, as claimed by the authors.

## 3. Modularity based community detection

Modularity based community detection happens to be another major type of community detection clustering algorithms in complex networks. Liu and Li [30] developed a novel metric representation, the co-neighbour modularity matrix, to assess the quality of community/clustering identification, by which the problem of community detection is transformed to that of a problem of clustering of eigenvectors in Euclidean space. Thereafter, the network community architecture is identified with spectral clustering algorithm. One major advantage of this algorithm is that it is free from the noise generated by the initial mean points of clustering e.g. in k-means category of algorithms.

Scibetta et al. [31] addressed the strategy of a division of the network into clusters or district metered areas for the detection of water losses from water distribution networks, since the measurement of incoming/outgoing flows for each cluster or district metered area allows for a quantification of water losses. The authors use the community detection approach developed in the complex network theory to identify clusters or district metered areas in water distribution systems. The method aims to find solutions satisfying the constraints of maximization of modularity and the reduction of the number of communities. The authors claim that the method is adequately scalable.

Sharma and Purohit [32] have also applied the spectral clustering algorithm for tracking community formation in complex social networks.

Zhuhadar et al. [33] have proposed the design of a visual recommender system to recommend learning resources to cyber learners within the same community, by using a community detection algorithm on the large scale complex networks of cyber learners and learning resources, based on Web Usage data of the subjects. Their algorithm uses a heuristic which initially accomplishes clustering by force-based visualization algorithm. Subsequently, the algorithm utilizes the information on network modularity to choose good decompositions from those found using visualization algorithm.

Yu and Ding [34] applied modularity clustering objective function for network community discovery. They have shown that a normalized form modularity clustering is equivalent to the prevalent normalized cut spectral clustering. They then use this information to interpret and solve the modularity clustering problem, and further corroborate the algorithm on some data collections.

## 4. Other algorithms

It is popularly accepted that social networks offer a dominant abstraction of the organization and dynamics of varied kinds of people or people-to-technology interaction, and also endorse the use of collaborative technologies for partnerships among different groups. It is also acknowledged that finding subgroups within social networks is important for understanding and possibly influencing the formation and evolution of online communities. In this context, Sharma and Joshi [35] address the issues of tracking online communities in large scale complex social networks. They infer the dynamics of the communities to a significant extent from the online interactions of the nodes by tracking the evolution of known sub-communities over time.

Verma and Butenko [36] use the clique relaxation concept of k-community for network community identification. The clique relaxation of k-community is a connected subgraph such that endpoints of each edge have at the minimum k shared neighbours within the subgraph. An important aspect of this method is that it does not use any previous information about the organization of the network. By defining a cluster as a k-community, the proposed algorithm aims to provide a clustering of a network into k-communities with varying values of k.

Lu et al. [37] proposed a novel Network Community Structure Clustering Algorithm Based on the Genetic Theory. Their work puts forward the idea of applying a clustering ensemble based genetic algorithm in the domain of complex social network mining. Their procedure introduces clustering ensemble into the crossover operator, and then utilizes the clustering information of the parents to generate new individuals. This seems to avoid the problems that are triggered by merely swapping strings between crossover operators without consideration of their contents. In population generation, Markov random walk approach is used to sustain the diversity of the entities as well as the clustering accuracy. The algorithm also uses a local searching mechanism in crossover operators to reduce the searching space and improve the speed of convergence.

Eagle et al. [38] demonstrate in what way using network community identifying methods can be used to recognize sub-goals in problems in a logic tutor, and then those community structures can be utilized to produce high level hints among sub-goals. The authors do this by presenting a new data structure, the Interaction Network, for representing interaction-data from open problem solving environment tutors.

## 5. Conclusion

The topology and evolution of real world complex networks are controlled by various organizing principles of topology and dynamics and happens to be a major research area. Researchers have been addressing structural and topological issues of complex networks as well as complex networks' dynamics for quite some time. Some of the important concepts that has evolved in the field of complex networks are small-world and scale-free networks, evolving networks, the relationship between topology and the network's robustness, degree distributions, clustering, network correlations, random graph models, models of network growth, dynamical processes on networks, etc. Researchers apply several types of methods and algorithms for detection and identification of communities in complex networks which can be generally divided into 4 categories. The optimal methods to detect network communities vary depending on applications, complex networks, computational resources etc. Some current lines of research on complex network communities are those on community evolution, overlapping communities, communities in directed networks, community characterization and interpretation, etc.

Several complex networks have the property of clustering or network transitivity. Many of the algorithms or methods proposed for network community detection through clustering are modified versions of or inspired from the concepts of minimum-cut based algorithms, hierarchical connectivity based algorithms, the original Girvan–Newman algorithm, concepts of modularity maximization, algorithms utilizing metrics from information and coding theory, and clique based algorithms. Consequently, Community Detection Clustering Algorithms in Complex Networks can be broadly classified into Hierarchy Based Algorithms and Information Theoretic Algorithms. Moreover there has been lot of research work carried out by eminent researchers in the fields of both Hierarchy Based Algorithms and Information Theoretic Algorithms as well as Modularity based community detection. Also, in case of social networks finding sub groups is important for understanding and possibly influencing the formation and evolution of online communities. Researchers have inferred the dynamics of the communities to a significant extent from the online interactions of the nodes by tracking the evolution of known sub-communities over time. Researchers have also used the clique relaxation concept of k-community for network community identification wherein by defining a cluster as a k-community the aim is to provide a clustering of a network into k-communities with varying values of k. Work has also been carried out based on genetic theory and a novel Network Community Structure Clustering Algorithm has been proposed based on it. Apart from these, significant research work has also been carried out based on network community identifying methods which can be used to recognize sub-goals in problems in a logic tutor, and then those community structures can be utilized to produce high level hints among sub-goals.

As can be seen community detection in complex networks is an active research area and has got real life applications in the fields of large scale engineering, social media analysis, biomedical data analysis, online education, business and economics etc. It is hoped that this survey will be of great help to the scientific community and senior researchers.

**References**

1. Albert, Réka, Hawoong Jeong, and Albert-László Barabási. "Error and attack tolerance of complex networks." *Nature* 406.6794 (2000): 378-382.

2. Albert, Réka, and Albert-László Barabási. "Statistical mechanics of complex networks." *Reviews of modern physics* 74.1 (2002): 47.

3. Milo, Ron, et al. "Network motifs: simple building blocks of complex networks." *Science* 298.5594 (2002): 824-827.

4. Strogatz, Steven H. "Exploring complex networks." *Nature* 410.6825 (2001): 268-276.

5. Stutzman, Frederic. "An evaluation of identity-sharing behavior in social network communities." *International Digital and Media Arts Journal* 3.1 (2006): 10-18.

6. Donetti, Luca, and Miguel A. Munoz. "Detecting network communities: a new systematic and efficient algorithm." *Journal of Statistical Mechanics: Theory and Experiment* 2004.10 (2004): P10012.

7. Fogel, Joshua, and Elham Nehmad. "Internet social network communities: Risk taking, trust, and privacy concerns." *Computers in Human Behavior* 25.1 (2009): 153-160.

8. Hattori, Fumio, et al. "Socialware: Multiagent systems for supporting network communities." *Communications of the ACM* 42.3 (1999): 55-ff.

9. Jain, Anil K., and Richard C. Dubes. *Algorithms for clustering data*. Prentice-Hall, Inc., 1988.

10. Hartigan, John A. "Clustering algorithms." (1975).

11. Mantel, Nathan. "The detection of disease clustering and a generalized regression approach." *Cancer research* 27.2 Part 1 (1967): 209-220.

12. Steinbach, Michael, George Karypis, and Vipin Kumar. "A comparison of document clustering techniques." *KDD workshop on text mining*. Vol. 400. No. 1. 2000.

13. Silva, Thiago C., and Liang Zhao. "Pixel clustering by using complex network community detection technique." Intelligent Systems Design and Applications, 2007. ISDA 2007. Seventh International Conference on. IEEE, 2007.

14. Zhang, Shihua, X-M. Ning, and X-S. Zhang. "Graph kernels, hierarchical clustering, and network community structure: experiments and comparative analysis." The European Physical Journal B 57.1 (2007): 67-74.

15. Xun, Guangxu, et al. "Latent Community Discovery with Network Regularization for Core Actors Clustering." COLING (Posters). 2012.

16. de Oliveira, Tatyana BS, et al. "Data clustering based on complex network community detection." Evolutionary Computation, 2008. CEC 2008.(IEEE World Congress on Computational Intelligence). IEEE Congress on. IEEE, 2008.

17. Cravino, Nuno, José Devezas, and Álvaro Figueira. "Using the overlapping community structure of a network of tags to improve text clustering." Proceedings of the 23rd ACM conference on Hypertext and social media. ACM, 2012.

18. Yang, Shuzhong, Siwei Luo, and Jianyu Li. "A novel visual clustering algorithm for finding community in complex network." Advanced Data Mining and Applications. Springer Berlin Heidelberg, 2006. 396-403.

19. Zhang, Shaozhong, et al. "Trust Network and Trust Community Clustering based on Shortest Path Analysis for E-commerce." International Journal of U-& E-Service, Science & Technology 5.2 (2012).

20. Zhang, Shaozhong, et al. "Trust Network and Small World Trust Community Clustering for E-Commerce." *International Journal of Hybrid Information Technology* 6.2 (2013): 1-14.

21. Piccardi, Carlo, Lisa Calatroni, and Fabio Bertoni. "Clustering financial time series by network community analysis." International Journal of Modern Physics C 22.01 (2011): 35-50.

22. Piccardi, Carlo, and Lisa Calatroni. "Clustering time series by network community analysis." Complexity in Engineering, 2010. COMPENG'10.. IEEE, 2010.

23. Huang, Hsun-Hui, and Horng-Chang Yang. "Semantic Clustering-Based Community Detection in an Evolving Social Network." Genetic and Evolutionary Computing (ICGEC), 2012 Sixth International Conference on. IEEE, 2012.

24. Liu, Zhi-Ping, et al. "Protein cavity clustering based on community structure of pocket similarity network." International journal of bioinformatics research and applications 4.4 (2008): 445-460.

25. Rui, Zhu, and Liu FengMing. "A clustering algorithm of community in distributed network based on trust." Fuzzy Systems and Knowledge Discovery (FSKD), 2011 Eighth International Conference on. Vol. 2. IEEE, 2011.

26. Xie, Jierui, and Boleslaw K. Szymanski. "Community detection using a neighborhood strength driven label propagation algorithm." Network Science Workshop (NSW), 2011 IEEE. IEEE, 2011.

27. Zhang, Shao-Zhong, et al. "Community clustering model for E-commerce trust based on social network." Journal of Zhejiang University. Engineering Science 47.4 (2013): 656-661.

28. Guan-yu, W. A. N. G. "Algorithm for Detecting Community of Complex Network Based on Clustering." Computer Engineering 10 (2011): 021.

29. OU, You-yuan, et al. "Web services clustering based on detecting community structure in complex network." Application Research of Computers 6 (2009): 090.

30. Liu, Ji, and Lei Li. "Network Community Detection Based on Co-Neighbor Modularity Matrix with Spectral Clustering." Applied Mechanics and Materials 55 (2011): 1237-1241.

31. Scibetta, Marco, et al. "Community detection as a tool for complex pipe network clustering." EPL (Europhysics Letters) 103.4 (2013): 48001.

32. Sharma, Sanjiv, And Gn Purohit. "A Novel Framework For Tracking Online Community Interaction In Social Network." International Journal Of Information Acquisition 9.02 (2013).

33. Zhuhadar, Leyla, Rong Yang, and Olfa Nasraoui. "Toward the Design of a Recommender System: Visual Clustering and Detecting Community Structure in a Web Usage Network." Proceedings of the The 2012 IEEE/WIC/ACM International Joint Conferences on Web Intelligence and Intelligent Agent Technology-Volume 01. IEEE Computer Society, 2012.

34. Yu, Linbin, and Chris Ding. "Network community discovery: solving modularity clustering via normalized cut." Proceedings of the Eighth Workshop on Mining and Learning with Graphs. ACM, 2010.

35. Sharma, Sanjiv, and N. K. Joshi. "Enhancement of existing clustering algorithm for tracking online community in social network."

36. Verma, Anurag, and Sergiy Butenko. "Network clustering via clique relaxations: A community based approach." Contemporary Mathematics 588 (2013).

37. Lu, Nan, Yuanyuan Jin, and Lei Qin. "Network Community Structure Clustering Algorithm Based on the Genetic Theory." Journal of Advances in Computer Networks 1.2 (2013).

38. [Eagle, Michael, Matthew Johnson, and Tiffany Barnes. "Interaction Networks: Generating High Level Hints Based on Network Community Clustering." International Educational Data Mining Society (2012).